\documentclass[fleqn,twoside]{article}
\usepackage{espcrc2}

\usepackage{graphicx,epsf}
\usepackage[figuresright]{rotating}

\title{M\"obius Fermions}

\author{R.C. Brower \address[MCSD]{Physics Department, 
                  590 Commonwealth Avenue,
                  Boston University Boston,
                  MA 02215, USA},
        H. Neff \address{Centre for Computational Science,
                  Chemistry Department, University College of London,
                  20 Gordon Street, London WC1H 0AJ, UK},
        K. Orginos \address[MCSD]{Department of Physics,
                  College of William and Mary,
                  Williamsburg, VA 23187-8795, USA}}

\begin{document}

\begin{abstract}
  We introduce a new domain wall operator that represents a full
  (real) M\"obius transformation of a given non-chiral Dirac kernel.
  Shamir's and Chiu/Bori\c{c}i's domain wall fermions are special cases of
  this new class. By tuning the parameters of the M\"obius operator
  and by introducing a new Red/Black preconditioning, we are able to
  reduce the computational effort substantially.
\end{abstract}

\maketitle


\section{Introduction}
\label{sec:intro}

The key idea in evading the Nielsen-Ninomiya no-go
theorem~\cite{Nielsen:1980rz,Nielsen:1981xu}, which forbade the
construction of lattice fermion actions with chiral symmetry under
rather general conditions, was introduced by
Kaplan~\cite{Kaplan:1992bt}. In his construction four dimensional
chiral zero modes appeared as bound states on a mass defect or 3-brane
in a five dimensional theory. Much like the work of Callan and
Harvey~\cite{Callan:1984sa} in the continuum, anomalous currents in
the 4 dimensional theory are understood as the flow on or off the mass
defect of conserved 5 dimensional currents.  This work led to two
concrete realizations of lattice fermions with chiral symmetry.  The
domain wall fermions~\cite{Shamir:1993zy,Furman:1994ky,Borici:1999zw,Chiu:2002ir} and the
overlap
fermions~\cite{Narayanan:1992wx,Narayanan:1993ss,Narayanan:1993sk,Neuberger:1997bg,Neuberger:1997fp}.

Here we introduce the M\"obius domain wall operator, a generalization
of Shamir's and Chiu/Bori\c{c}i's suggestions. It is given by (to keep the
notation simple, we choose the length of the fifth domain wall
dimension $L_s$ equal to 4)
\begin{eqnarray}\label{eq:mobius}
D_{DW}(m)= \hskip2cm \\
 \left(\begin{array}{cccc}
\hskip-.13cm D^{(1)}_+  & \hskip-.13cm D^{(1)}_- P_- & \hskip-.13cm  0  & \hskip-.13cm   -mD^{(1)}_- P_+ \\
\hskip-.13cm D^{(2)}_- P_+  & \hskip-.13cm  D^{(2)}_+  & \hskip-.13cm  D^{(2)}_- P_-  & \hskip-.13cm  0  \\
\hskip-.13cm 0  & \hskip-.13cm  D^{(3)}_- P_+  & \hskip-.13cm  D^{(3)}_+  & \hskip-.13cm  D^{(3)}_- P_-  \\
\hskip-.13cm -mD^{(4)}_- P_-  & \hskip-.13cm  0  & \hskip-.13cm  D^{(4)}_- P_+  & \hskip-.13cm  D^{(4)}_+  \\
\end{array} \hskip-.13cm \right)\hskip-.38cm \nonumber
\end{eqnarray}
with
\begin{eqnarray}
D^{(i)}_+ &=& b_i D_w +1 , \;\;\; D^{(i)}_- = c_i D_w -1 \label{eq:coeff} ,\\
P_+ &=& \frac{1}{2} ( 1 + \gamma_5) , \;\;\; P_- = \frac{1}{2} ( 1 - \gamma_5) .
\end{eqnarray}
$D_w$ denotes the Wilson Dirac matrix
\begin{eqnarray}\label{eq:wilson}
D_w (M_5) =  (4+M_5) \delta_{x,y} - \frac{1}{2}  \bigl[  (1  -  \gamma_\mu) \\
 U_\mu(x) \delta_{x+\mu,y}  
+  (1  +   \gamma_\mu) U_\mu^\dagger(y) \delta_{x,y+\mu}  \bigr] .\nonumber
\end{eqnarray}
Note, that this choice is not mandatory, but that any other Dirac
operator  could have been used here as well.

Eq.(\ref{eq:mobius}) is a generic expression for the domain wall
fermions. The different operators, Shamir, Chiu/Bori\c{c}i and M\"obius,
are characterized by the coefficients $b_i, c_i$ in
eq.(\ref{eq:coeff}). The M\"obius operator contains Shamir's and
Chiu/Bori\c{c}i's suggestions as special cases. For M\"obius, the
coefficients are only constrained to be:
\begin{itemize}
\item $ b_i, c_i \; \in  \;\Re $  ,
\item $b_i-c_i=const, \;\forall\; i \;\in \; L_s$ (i.e.\ $const$ is independent of $i$).
\end{itemize}
Shamir and Chiu/Bori\c{c}i use:
\begin{eqnarray}
\begin{array}{ccc}
\mbox{Shamir:} & b_i=a,  & c_i=0  ,\\
\mbox{Bori\c{c}i:} & b_i=a,  & c_i=a ,\\
\mbox{Chiu:} & b_i=a_i,  & c_i=a_i  .
\end{array} 
\end{eqnarray}
with $a, a_i \; \in  \;\Re . $

To understand the meaning of the coefficients, we will translate the 5
dimensional domain fermions into a 4 dimensional overlap
operator. This is done via a linear matrix transformation.

\section{Domain wall - overlap transformation}

The domain wall and the overlap operator are connected through a
linear matrix transformation
\cite{Kikukawa:1999sy,Edwards:2000qv,Brower:1997ha}. The length of the
fifth domain wall dimension corresponds to the order of a polynomial,
that approximates the sign function on the overlap side. Accordingly,
domain wall fermions can be seen as a preconditioning of the overlap
operator.

In the following, $D_{DW}$ will denote the generic domain wall operator
and $L_s$ the length of the fifth domain wall dimension. Here we will
choose $L_s=4$ to keep the notation simple, but all formulas hold for
any $L_s$. $D_{OV}$ will denote the approximation to the overlap
operator, as defined through the polynomial of finite order, that
describes the sign function.

The domain wall - overlap transformation  reads:
\begin{equation}\label{eq:4d-5d}
L \;  D_{DW}(m) \; R \; = \; F \; D^5_{OV}(m),
\end{equation}
with
\begin{equation}
F \; = L \;  D_{DW}(1) \; R ,
\end{equation}
\begin{eqnarray}
L= 
\left(\begin{array}{cccc}
1 & S_1   & S_1 S_2    & S_1 S_2 S_3   \\
0 & 1        & S_2           & S_2 S_3   \\
0 & 0        & 1                & S_3   \\
0 & 0        & 0                & 1 
\end{array} \right) \times
\end{eqnarray}
\begin{eqnarray}
\left(\begin{array}{cccc}
{Q_-^{(1)}}^{-1}  & 0  & 0 & 0  \\
0 & {Q_-^{(2)}}^{-1}    & 0 & 0  \\
0 & 0  & {Q_-^{(3)}}^{-1}   & 0  \\
0 & 0  & 0 & {Q_-^{(4)}}^{-1}   \\
\end{array} \right)\gamma_5  , \nonumber
\end{eqnarray}
\begin{eqnarray}
R = P\hat{R} =\hskip2.3cm\\
\left(\begin{array}{cccc}
\hskip-.13cm P_- & \hskip-.13cm  P_+  & \hskip-.13cm  0 & \hskip-.13cm  0  \\
\hskip-.13cm 0 & \hskip-.13cm  P_-  & \hskip-.13cm  P_+ & \hskip-.13cm  0  \\
\hskip-.13cm 0 & \hskip-.13cm  0  & \hskip-.13cm  P_- & \hskip-.13cm  P_+  \\
\hskip-.13cm P_+ & \hskip-.13cm  0  & \hskip-.13cm  0 & \hskip-.13cm  P_- \\
\end{array} \right) \hskip-.13cm
\left(\begin{array}{cccc}
\hskip-.13cm  -1 & \hskip-.13cm  0  & \hskip-.13cm  0 & \hskip-.13cm  0  \\
\hskip-.13cm  -S_2 S_3 S_4 \, d & \hskip-.13cm  1  & \hskip-.13cm  0 & \hskip-.13cm  0  \\
\hskip-.13cm  -S_3 S_4 \, d & \hskip-.13cm  0  & \hskip-.13cm  1 & \hskip-.13cm  0  \\
\hskip-.13cm  -S_4 \, d & \hskip-.13cm  0  & \hskip-.13cm  0 & \hskip-.13cm  1 \\
\end{array} \right),\hskip-.38cm \nonumber
\end{eqnarray}
\begin{eqnarray}
 D^5_{OV}(m)= \left(\begin{array}{cccc}
D_{OV}(m) & 0  & 0 & 0  \\
0 & 1  & 0 & 0  \\
0 & 0  & 1 & 0  \\
0 & 0  & 0 & 1 \\
\end{array} \right).
\end{eqnarray}

The matrix entries are defined as follows:

\begin{eqnarray}
Q^{(i)}_+ &=& \gamma_5 D_w (b_i P_+ + c_i P_-) +1, \\
Q^{(i)}_- &=& \gamma_5 D_w (b_i P_- + c_i P_+) -1,\\
S_i &=& T_i^{-1} = - (Q_-^{(i)})^{-1} Q_+^{(i)} ,\\
d &=& P_+ - m P_- . 
\end{eqnarray}

$T_i^{-1}$ is called the transfer matrix. 

Multiplying the matrices on the left hand side of eq.(\ref{eq:4d-5d})
(multiply first by $P$, then $L$ and $\hat{R}$, where it might be
useful to remember that $(b P_- + c P_+ )^{-1} = \frac{1}{b} P_- +
\frac{1}{c} P_+ $) leads to the entry $[L D_{DW} (m) R ]_{11} = -(P_- - mP_+) +
S (P_+ -mP_-)$, or
\begin{eqnarray}
 F= \left(\begin{array}{cccc}
(S+1) \gamma_5 & 0  & 0 & 0  \\
0 & 1  & 0 & 0  \\
0 & 0  & 1 & 0  \\
0 & 0  & 0 & 1 \\
\end{array} \right) 
\end{eqnarray}
and
\begin{equation}\label{eq:overlap}
D_{OV}(m)=\frac{1}{2} \left(
1+m + (1-m) \gamma_5 \frac{(S-1)}{(S+1)}
\right),
\end{equation}
with $ S=S_1S_2S_3S_4$ (note, that the fact that there are four factors
$S_1, \cdots, S_4$ is due to our choice $L_s=4$).  If
${(S-1)}/{(S+1)}$ was an approximation to the sign function,
eq.(\ref{eq:overlap}) would be the corresponding approximation to the
overlap operator.  To see whether there is such a relation, we define
$H^{(i)}_T$ through:
\begin{equation}\label{eq:kernel}
S_i=\frac{H^{(i)}_T+1}{H^{(i)}_T-1},
\end{equation}
i.e.\
\begin{equation}\label{eq:scaling}
H^{(i)}_T=(b_i+c_i)\gamma_5 D_w \frac{1}{2+(b_i-c_i)D_w}.
\end{equation}
We define the kernel $H_T$ as
\begin{equation}
H_T=\gamma_5 D_w \frac{1}{2+(b_i-c_i)D_w},
\end{equation}
i.e.\
\begin{equation}
H^{(i)}_T=(b_i+c_i) H_T.
\end{equation}
This leads to
\begin{equation}\label{eq:poldecomp}
\frac{(S-1)}{(S+1)}=\frac{A-B}{A+B},
\end{equation}
with
\begin{eqnarray}\label{eq:poldecomp1}
A=(H^{(1)}_T \hskip-.05cm + \hskip-.05cm 1)(H^{(2)}_T \hskip-.05cm + \hskip-.05cm 1)(H^{(3)}_T \hskip-.05cm + \hskip-.05cm 1)(H^{(4)}_T \hskip-.05cm + \hskip-.05cm 1),\\
B=(H^{(1)}_T \hskip-.05cm - \hskip-.05cm 1)(H^{(2)}_T \hskip-.05cm - \hskip-.05cm 1)(H^{(3)}_T \hskip-.05cm - \hskip-.05cm 1)(H^{(4)}_T \hskip-.05cm - \hskip-.05cm 1).\label{eq:poldecomp2}
\end{eqnarray}
One can choose the coefficients $b_i$ and $c_i$ such that
eq.(\ref{eq:poldecomp}) corresponds to an approximation $\epsilon$ to
the sign function with kernel $H_T$.  As mentioned above, we set
$b_i-c_i $ equal to a constant value for all $i$, i.e.\ the
denominator is independent of $i$.

Possible polynomial approximations are:
\begin{itemize}
\item $c_i+b_i=const , \;\forall\; i \;\in \; L_s $. This corresponds
to Neuberger's polar decomposition.
\item $c_i+b_i$ equal to Zolotarev's coefficients \cite{Chiu:2002ir}.
\end{itemize}

We can summarize this findings as
\begin{equation}\label{eq:overlap1}
D_{OV}(m)=\frac{1}{2} \left(1+m + (1-m)\; \gamma_5 \; \epsilon (H_T) \right).
\end{equation}

\section{Domain Wall preconditioning}

Here we will describe how the quark propagator can be determined via
the 5 dimensional domain wall operator.  

We are interested in the 4 dimensional propagator $x_1$ (corresponding
to the source b), given by

\begin{equation}\label{eq:4d-equation}
D_{OV} x_1 = b .
\end{equation}

Eq.(\ref{eq:4d-5d}) can be used to precondition
eq.(\ref{eq:4d-equation}).  As it stands, eq.(\ref{eq:4d-5d}) is not
suitable for this task. We therefore perform the following
simplifications:
\begin{itemize}
\item  Multiply eq.(\ref{eq:4d-5d}) by $F^{-1}$ from the left
\begin{equation}
\hat{R}^{-1} P^{-1} D^{-1}_{DW} (1) D_{DW}(m) P\hat{R}  = D_{OV} ,
\end{equation}
\item  then multiply by $\hat{R}$ from the left and by $\hat{R}^{-1}$ from the  right
\begin{equation}\label{eq:simplify}
P^{-1} D^{-1}_{DW} (1) D_{DW}(m) P  = \hat{R}D_{OV}\hat{R}^{-1} .
\end{equation}
\end{itemize}
The right hand side of eq.(\ref{eq:simplify}) is then given by (with $d= P_+ - m P_-$)
\begin{eqnarray}
\hat{R}D_{OV}\hat{R}^{-1}\hskip-.14cm =\hskip-.14cm  \left(\begin{array}{cccc}
\hskip-.14cm  D_{OV} &  \hskip-.14cm   0  &\hskip-.14cm    0 & \hskip-.14cm   0  \\
\hskip-.14cm  S_2 S_3 S_4 (D_{OV} \hskip-.1cm - \hskip-.1cm 1) \, d & \hskip-.14cm  1  & \hskip-.14cm   0 &\hskip-.14cm    0  \\
\hskip-.14cm  S_3 S_4  (D_{OV} \hskip-.1cm -\hskip-.1cm 1) \, d & \hskip-.14cm   0  &\hskip-.14cm   1 &\hskip-.14cm    0  \\
\hskip-.14cm  S_4  (D_{OV} \hskip-.1cm  -\hskip-.1cm  1) \, d &  \hskip-.14cm 0  &\hskip-.14cm    0 & \hskip-.14cm   1 \\
\end{array} \right) . \hskip-.38cm 
\end{eqnarray}
Obviously, with 
\begin{eqnarray}
\left(\begin{array}{cccc}
\hskip-.14cm  D_{OV} &  \hskip-.14cm   0  &\hskip-.14cm    0 & \hskip-.14cm   0  \\
\hskip-.14cm  S_2 S_3 S_4 (D_{OV} \hskip-.1cm - \hskip-.1cm 1) \, d & \hskip-.14cm  1  & \hskip-.14cm   0 &\hskip-.14cm    0  \\
\hskip-.14cm  S_3 S_4  (D_{OV} \hskip-.1cm -\hskip-.1cm 1) \, d & \hskip-.14cm   0  &\hskip-.14cm   1 &\hskip-.14cm    0  \\
\hskip-.14cm  S_4  (D_{OV} \hskip-.1cm  -\hskip-.1cm  1) \, d &  \hskip-.14cm 0  &\hskip-.14cm    0 & \hskip-.14cm   1 \\
\end{array} \right)\hskip-.20cm 
\left(\begin{array}{c}
\hskip-.14cm  x_1 \hskip-.14cm  \\
\hskip-.14cm  x_2 \hskip-.14cm  \\
\hskip-.14cm  x_3 \hskip-.14cm  \\
\hskip-.14cm  x_4 \hskip-.14cm  \\
\end{array} \right)
\hskip-.14cm =\hskip-.14cm 
\left(\begin{array}{c}
\hskip-.14cm  b \hskip-.14cm  \\
\hskip-.14cm  0 \hskip-.14cm  \\
\hskip-.14cm  0 \hskip-.14cm  \\
\hskip-.14cm  0 \hskip-.14cm  \\
\end{array} \right)
\end{eqnarray}
$x_1$ is still the solution of $D_{OV} x_1 = b$.  Equivalently, we can
find $x_1$ by solving the left hand side of eq.(\ref{eq:simplify})
\begin{eqnarray}
P^{-1} D^{-1}_{DW} (1) D_{DW}(m) P  
\left(\begin{array}{c}
 x_1  \\
 x_2  \\
 x_3  \\
 x_4  \\
\end{array} \right)
=
\left(\begin{array}{c}
 b  \\
 0  \\
 0  \\
 0  \\
\end{array} \right)
\end{eqnarray}
or
\begin{eqnarray}
 D_{DW}(m) P  
\left(\begin{array}{c}
 x_1  \\
 x_2  \\
 x_3  \\
 x_4  \\
\end{array} \right)
=  D_{DW} (1) P
\left(\begin{array}{c}
 b  \\
 0  \\
 0  \\
 0  \\
\end{array} \right) .
\end{eqnarray}
Typically, in linear system solvers, one iterates $D_{DW}(m) \vec{y} =
\vec{b}$ alone, i.e.\ without the matrix $P$. Therefore, one has to
reconstruct the real solution $x_1$ as 

\begin{equation}
x_1 = P_- y_1 + P_+ y_4 .
\end{equation}

This follows directly from $D_{DW} = D_{DW} P P^{-1}$, i.e\ $\vec{x} = P^{-1} \vec{y}$ .

\section{The sign function}

For later reference, we will state here a few simple properties of the sign function.

The sign function satisfies the following equation
\begin{equation}\label{eq:sign1}
\mbox{sign} (x) = \mbox{sign} (\lambda x), \; \forall \; x \;\in \;\Re \;,\; \lambda \; \in \;\Re^+ .
\end{equation}

\begin{figure}[t]
\hskip-.6cm  \includegraphics[angle=270,width=18pc]{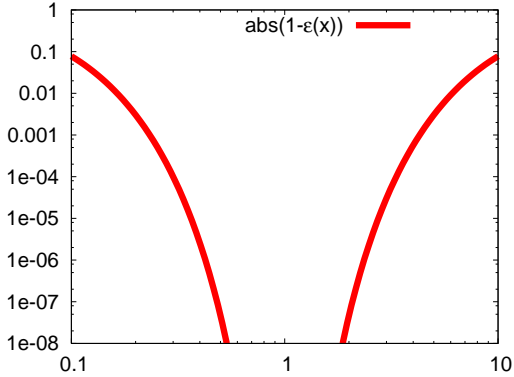}
\caption{$\epsilon(x)$is given by the polar decomposition of order 16:
$\epsilon(x) = \frac{(1+x)^{16}-(1-x)^{16}}{(1+x)^{16}+(1-x)^{16}}$.}
\label{fig:neuberger}
\vskip-.174cm
\end{figure}

Let $\epsilon$ be a polynomial approximation to the sign function. For
$\epsilon$, eq.(\ref{eq:sign1}) doesn't hold, instead we have

\begin{equation}\label{eq:sign2}
\epsilon (x)\neq  \epsilon (\lambda x) .
\end{equation}
Eq.(\ref{eq:sign2}) can easily be understood by looking at
fig.(\ref{fig:neuberger}). There, we show the quality of the polar
decomposition, by plotting its deviation from the sign
function. Obviously, the approximation is best for $x \approx 1$. The factor
$\lambda$ slides this curve along the abscissa.

For the overlap operator (eq.(\ref{eq:overlap})) this means:
\begin{itemize}
\item Eq.(\ref{eq:sign1}) states that scaling the kernel $H_T$ is a valid operation.
\item Eq.(\ref{eq:sign2}) demonstrates that the quality of the
approximation to the sign function depends on this scaling. In
other words, there is an optimal scaling factor.
\end{itemize}

\section{Comparison of the different domain wall fermions}

As demonstrated earlier, the kernels of the domain wall
actions are given by (see eq.(\ref{eq:kernel}))
\begin{eqnarray}\label{eq:kernels}
&\mbox{M\"obius:} \; (b_i+c_i) \gamma_5 D_w  (2+(b_i-c_i) D_w)^{-1}& \\
&\mbox{Shamir:}\;  a\gamma_5 D_w  (2+a D_w)^{-1}  , & \\
&\mbox{Bori\c{c}i:}\;  a \gamma_5 D_w , \;\;\; \mbox{Chiu:} \; a_i \gamma_5 D_w .&
\end{eqnarray}

In the following, we will show how the determination of the quark
propagator depends on the choice of the coefficients $b_i, c_i$ (and
hence on $a, a_i$ as well). We will describe the advantages of the M\"obius
operator as compared to Shamir's and Chiu/Bori\c{c}i's suggestions.

\subsubsection{The M\"obius operator}

As can be seen in eq.(\ref{eq:kernels}), the M\"obius kernel can be
scaled with the coefficient $b_i+c_i$. In other words, the eigenvalues
of the operator $\gamma_5 D_w (2+(b_i-c_i) D_w)^{-1}$ can be slided
along the abscissa until the approximation to the sign function is
optimal (with given $L_s$).

The coefficients $b_i - c_i$ in the denominator are different. They
don't simply act as scaling factors, but change the spectrum of the
operator.  In other words, for each $b_i - c_i$ one has a different
matrix. Accordingly, one can use $b_i - c_i$ to tune the condition
number of the kernel operator. The smaller the condition number, the
better the approximation to the sign function will be.

As mentioned above, the denominator will always be chosen independent
of $i$. Not so for $b_i+c_i$. We have $L_s$ different
coefficients. This freedom allows for different choices of polynomials
that approximate the sign function on the overlap side, such as the
polar decomposition or Zolotarev's polynomials (see
eq.(\ref{eq:poldecomp1}) and eq.(\ref{eq:poldecomp2})).

\subsubsection{Shamir's operator}

Shamir's operator allows for a tuning of the condition number, since
it possesses a coefficient $a$ in the denominator. On the other hand,
the same coefficient $a$ acts as the scaling factor in the
numerator. Therefore this operator cannot be scaled without changing
the matrix itself. In other words, in the two dimensional space,
spanned by the coefficients $b_i+c_i$ and $b_i-c_i$, Shamir's operator
can only exploit the diagonal.

It follows that in this case the only possible polynomial
approximation on the overlap side is Neuberger's polar decomposition.

\subsubsection{Chiu/Bori\c{c}i's operator}

Chiu/Bori\c{c}i's action has independent coefficients $a_i$ that act solely as
scaling factors. On the other hand, the denominator is constant (equal
to 2). Therefore the condition number for Chiu/Bori\c{c}i's operator can not be
tuned. Note, that this operator correspond to the standard overlap
approach, which employs Dirac fermions, with a denominator
equal to the identity.

\subsubsection{Conclusions}

M\"obius fermions are a best of two worlds approach. They combine
Shamir's tuning of the condition number with the scalability of
Chiu/Bori\c{c}i's action. Our results will demonstrate that this leads to a
significant reduction of the computational costs.

\section{Red/black preconditioning}

The standard red/black preconditioning is only applicable for Shamir's
action.  We therefore introduce a new red/black partitioning that can
be employed for the M\"obius operator (and hence for Shamir and Chiu/Bori\c{c}i
as well).

The two preconditioning methods are defined as follows:
\begin{itemize}
\item Standard red/black: every neighbour of a black point is red.
\item New red/black: every space-time neighbour of a black point is
red, every neighbour in the fifth dimension of a black point is
black.
\end{itemize}

A matrix $M$, acting on a vector $x$, can then be written as:
\begin{eqnarray}
M x =\left(\begin{array}{cc}
M_{rr} & M_{rb} \\
M_{br} & M_{bb} 
\end{array}    \right) 
\left(\begin{array}{c}
x_{r}\\
x_{b} 
\end{array}    \right) .
\end{eqnarray}
Red/black preconditioning of this matrix is then defined through
\begin{eqnarray}
L M R = \left(\begin{array}{cc}
\hskip-.15cm M_{rr} & \hskip-.15cm 0 \\
\hskip-.15cm 0 & \hskip-.15cm M_{bb} - M_{br} M^{-1}_{rr} M_{rb} 
\end{array}   \hskip-.15cm  \right) ,
\end{eqnarray}
with
\begin{eqnarray}
L  \hskip-.09cm =  \hskip-.09cm \left(\begin{array}{cc}
\hskip-.15cm I_{rr} &\hskip-.15cm  0 \\
\hskip-.15cm -M_{br} M^{-1}_{rr} &\hskip-.15cm  I_{bb} 
\end{array}  \hskip-.15cm   \right),
R  \hskip-.1cm =  \hskip-.1cm \left(\begin{array}{cc}
\hskip-.15cm I_{rr} & \hskip-.15cm - M^{-1}_{rr} M_{rb} \\
\hskip-.15cm 0  & \hskip-.15cm I_{bb} 
\end{array}   \hskip-.15cm  \right)  .  \hskip-1.15cm\nonumber
\end{eqnarray}

$I$ is the identity operator.  In principle, the two sets (red and
black) can be chosen freely. But from a practical point
of view, $M^{-1}_{rr}$ has be a simple matrix, since it has to be
inverted in each iteration step of the linear system solver.

To keep notation simple, let's define for the Wilson operator
(eq.(\ref{eq:wilson})) $\hat{m}, \hat{D}$ as $D_w (M_5) = \hat{m}
\delta_{x,y} - 1/2 \hat{D}$. For standard red/black
preconditioning we find
\begin{eqnarray}
M^{standard}_{rr}= \hskip2.7cm  \\
\left(\hskip-.1cm \begin{array}{cccc}
& \cdots &&\\
-\frac{1}{2} \hat{D}^{(2)} P_+ &   b_2  \hat{m}  +1
& -\frac{1}{2} \hat{D}^{(2)} P_- & 0 \\
& \cdots &&\\
& \cdots &&\\
\end{array} \hskip-.1cm  \right)  .   \nonumber
\end{eqnarray}
This matrix is computationally too costly, due to the off diagonal terms
$c_i \hat{D}$ (note that for Shamir's operator $c_i = 0 $).

For the new preconditioning method, on the other hand, we find
\begin{eqnarray}
M^{new}_{rr}= \hskip2.7cm  \\
\left(\hskip-.1cm  \begin{array}{cccc}
&\hskip-.1cm \cdots &&\\
  (c_2  \hat{m}  -1) P_+ & \hskip-.1cm  b_2  \hat{m}  +1
& \hskip-.1cm  (c_2  \hat{m}  -1) P_-  & \hskip-.1cm 0 \\
& \cdots &&\\
& \cdots &&\\
\end{array}  \hskip-.1cm  \right) . \hskip-1.1cm \nonumber
\end{eqnarray}
$M^{new}_{rr}$ only contains coefficients and the chiral projectors
$P_{\pm}$ and thus can be inverted analytically. Therefore, its cost in
the linear system solver is negligible.

The standard preconditioning, which can be applied to Shamir's
operator, results in a numerical speed up of roughly $2.6$. We find
the same acceleration for the M\"obius operator with the new
preconditioning method. Both methods are therefore equivalent in terms
of convergence, whereas only the new approach is generally applicable.

\section{Results}

We will present our results for the M\"obius operator. As our measure
of performance we count the number of Wilson Dirac applications that
the linear system solver, the conjugate gradient method on the normal
equation $D_{DW}^{\dagger} D_{DW}$, needs to converge (note, that even
though the M\"obius operator contains three Dirac matrices per row,
see (eq.(\ref{eq:mobius})), whereas Shamir only one, both operators
only require $L_s$ Wilson Dirac applications per $D_{DW}$
application).  The quality of the approximation to the sign function
is measured via the residual mass \cite{Aoki:2002vt,AliKhan:2000iv,Blum:2000kn}. 
We will find that M\"obius' more
general set of coefficients, as compared to Shamir and Chiu/Bori\c{c}i,
leads to a substantial reduction of the computational effort.

We perform our measurements on 20 quenched $16^3 \times 32$ 
gauge fields, generated with the Wilson action at $\beta=6.0$.

Our point of reference will be Shamir's operator with a widely used
set of parameters, $M_5=1.8$ and $b_i-c_i=1.0$. We will refer to this
setting as 'standard Shamir'.

For each set of parameters, we have to tune the quark mass $m$, such
that the pion mass agrees with standard Shamir.  Since we find the
residual mass of M\"obius with $L_s=8$ to be roughly equal to standard
Shamir with $L_s=16$, we adjust the pion mass such that it agrees for
this two cases. The pion mass dependence on the scaling factor
$b_i+c_i$ is weak and will therefore be neglected.

In all graphs, the 'number of Dirac applications' is normalized
such that it represents $L_s $ times the number of iterations per
source. Hence, we neglect a factor two, due to the normal
equation, $D_{DW}^{\dagger} D_{DW}$.

\begin{figure}[tb]
\hskip-.5cm  \includegraphics[angle=270,width=19pc]{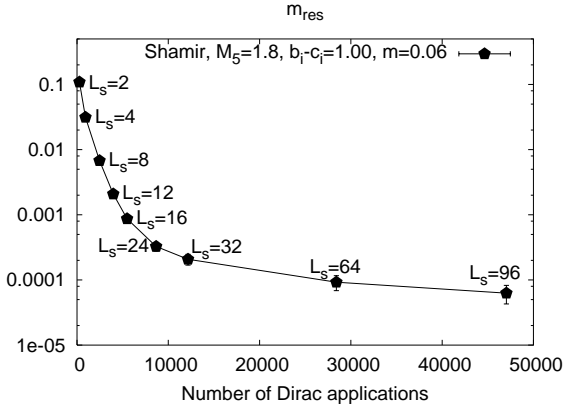}
\caption{'Standard Shamir' with $M_5=1.8$ and $b_i-c_i=1.0$}
\label{fig:Fig0}
\vskip-.174cm
\end{figure}

\subsection{Figure \ref{fig:Fig0}}

Fig.(\ref{fig:Fig0}) shows the $L_s$ dependence of the residual mass for
standard Shamir.  We will compare all our results to this data.

\begin{figure}[t]
\hskip-.5cm \includegraphics[angle=270,width=19pc]{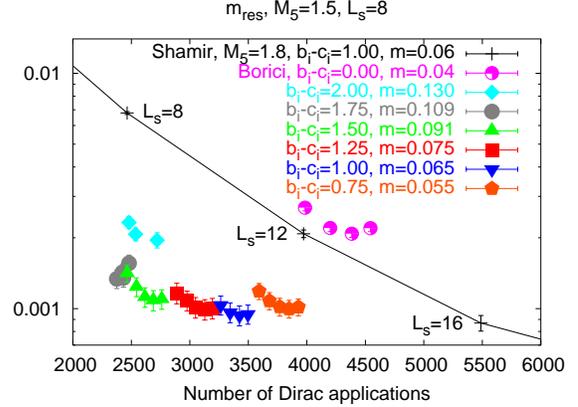}
\caption{Comparison of the M\"obius operator with standard Shamir at
$m_q=0.06$ and $M_5=1.5$ (except Shamir, which uses $M_5=1.8$). The
series of points for a given $b_i-c_i$ corresponds to different values
of $b_i+c_i$. Neighbouring points represent $b_i+c_i$ values which
differ by 0.1. The $b_i+c_i$ values are, $[$Bori\c{c}i: $b_i+c_i=1.0,
\ldots,1.3 ]$, $[b_i-c_i=0.75: b_i+c_i=2.2, \ldots, 2.6 ]$,
$[b_i-c_i=1.0: b_i+c_i=2.2,\ldots, 2.5 ]$, $[b_i-c_i=1.25:
b_i+c_i=2.0, \ldots, 2.4 ]$, , $[b_i-c_i=1.5:b_i+c_i=1.8, \ldots, 2.2
]$, $[b_i-c_i=1.75: b_i+c_i=1.7, \ldots, 2.0]$, $[b_i-c_i=2.0:
b_i+c_i=1.6, 1.9, 2.0 ]$. The optimal $b_i-c_i$ values are:
Chiu/Bori\c{c}i: $b_i+c_i=1.2$, $b_i-c_i=0.75: b_i+c_i=2.5$, $b_i-c_i=1.0:
b_i+c_i=2.4 $, $b_i-c_i=1.25: b_i+c_i=2.3$, ,
$b_i-c_i=1.5:b_i+c_i=2.1$, $b_i-c_i=1.75: b_i+c_i=1.9$, $b_i-c_i=2.0:
b_i+c_i=1.6$. }
\label{fig:Fig1}
\vskip-.174cm
\end{figure}

\subsection{Figure \ref{fig:Fig1}}

In fig.(\ref{fig:Fig1}), we present our results for the M\"obius
operator at $L_s=8$ and $M_5=1.5$, but various $b_i-c_i$. We use a
Shamir quark mass of $m=0.06$, which corresponds to a pion mass in
lattice units of $m_{\pi}=0.44$. The series of points, for a given
$b_i-c_i$, corresponds to different values of $b_i+c_i$. It is important
to note that we choose all scaling coefficients to be the same, i.e.\
$b_1+c_1=b_i+c_i, \;\forall\; i \;\in \; L_s$. In other words, we
choose the polar decomposition to approximate the sign function.

In general, the number of Dirac applications increases with $b_i+c_i$. 
For $b_i+c_i=1.75$ this behaviour starts to change. At
$b_i+c_i=2.0$ it is even reversed and the number of Dirac
applications falls with growing $b_i+c_i$.

\begin{figure}[t]
\hskip-.5cm \includegraphics[angle=270,width=19pc]{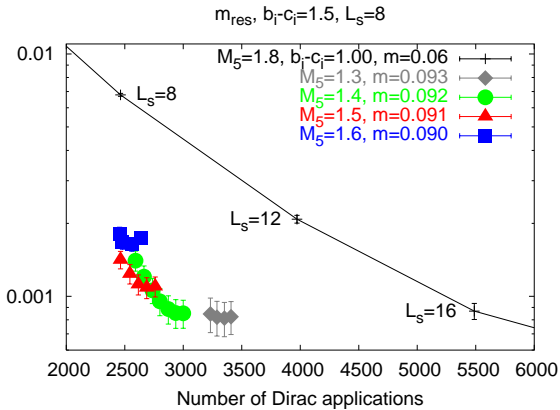}
\caption{$M_5$ dependence of the M\"obius operator, with optimal
$b_i-c_i$ as determined in fig.(\ref{fig:Fig1}).  $M_5=1.4$ and
$M_5=1.5$ are optimal, where $M_5=1.4$ reaches smaller residual
masses, but $M_5$ requires less Wilson Dirac applications. The
$b_i+c_i$ values are, $[M_5=1.3: b_i+c_i=2.9, \ldots, 3.2]$,
$[M_5=1.4: b_i+c_i=2.0, \ldots, 2.6]$, $[M_5=1.5: b_i+c_i=1.8, \ldots,
2.2]$, , $[M_5=1.6: b_i+c_i=1.6, \ldots, 1.9]$.}
\label{fig:Fig3}
\vskip-.174cm
\end{figure}

\subsection{Figure \ref{fig:Fig3}}

In fig.(\ref{fig:Fig3}), we analyse the dependence of M\"obius'
residual mass on $M_5$, see eq.(\ref{eq:wilson}). We use the optimal
$b_i-c_i=1.5$, as determined in the analysis in
fig.(\ref{fig:Fig1}). Again, the number of Wilson Dirac applications
increases with $b_i+c_i$. We find that the optimal $M_5$ values are
$M_5=1.4$ and $M_5=1.5$, where $M_5=1.4$ reaches smaller residual
masses, but $M_5$ requires less Wilson Dirac applications.

As can be read off from the abscissa, for the optimal $b_i-c_i$ and
$M_5$ values, the M\"obius operator is roughly two times cheaper
than standard Shamir.

\begin{figure}[t]
\hskip-.5cm  \includegraphics[angle=270,width=19pc]{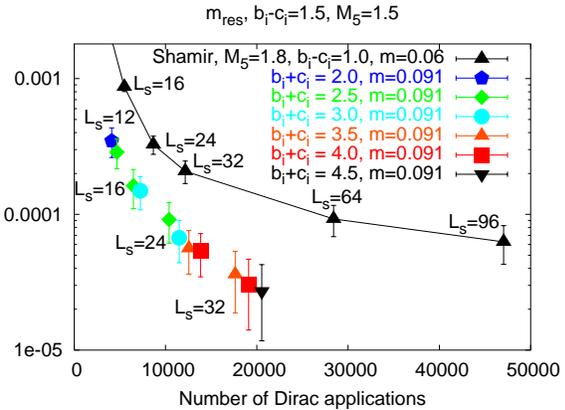}
\caption{$L_s$ dependence of the residual mass, with $b_i-c_i=1.5$ and
$M_5=1.5$. It can be seen that M\"obius is numerically much better
behaved than standard Shamir.}
\label{fig:Fig4}
\vskip-.174cm
\end{figure}

\subsection{Figure \ref{fig:Fig4}}

In fig.(\ref{fig:Fig4}), we show how M\"obius' residual mass behaves
for larger $L_s$, with $b_i-c_i=1.5$ and $M_5=1.5$. Obviously, the
relative improvement over standard Shamir grows rapidly.

\begin{figure}[t]
\hskip-.5cm  \includegraphics[angle=270,width=19pc]{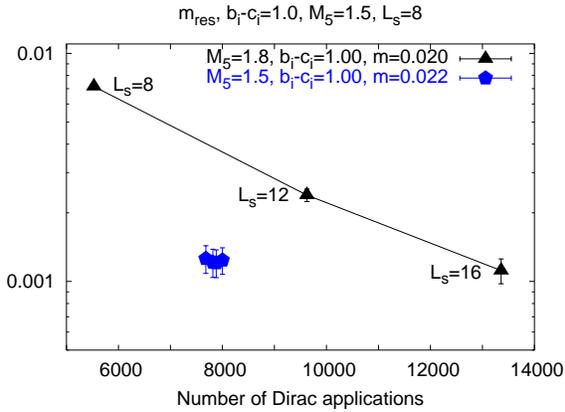}
\caption{Residual mass, for smaller standard Shamir quark mass
$m=0.02$, with $M_5=1.5$, $b_i-c_i=1.0$ and $b_i+c_i=2.3, \ldots,
2.6$. The factor of improvement over standard Shamir grows from 1.6 to
1.7, as compare to the $m=0.06$ analysis. The optimal $b_i+c_i$ values
is again 2.4, as for $m=0.06$.}
\label{fig:Fig5}
\vskip-.174cm
\end{figure}

\subsection{Figure \ref{fig:Fig5}}

In fig.(\ref{fig:Fig5}), we consider the behaviour of M\"obius'
residual mass for the smaller standard Shamir quark mass $m=0.02$,
with $M_5=1.5$ and $b_i-c_i=1.0$. Two facts are worth
mentioning. Firstly, compared to the analysis with $m=0.06$, the
factor of improvement for these $M_5$ and $b_i-c_i$ values, increases
from 1.6 to 1.7. This means that the advantage of M\"obius over
standard Shamir grows with falling quark mass. Secondly, the optimal
$b_i+c_i$ is equal to 2.4, both for $m=0.02$ and $m=0.06$. This
suggest, that the tuning of the M\"obius operator can be performed at
heavy quark masses, where the computation of the propagators is less
expensive.

\begin{figure}[t]
\hskip-.5cm  \includegraphics[angle=270,width=19pc]{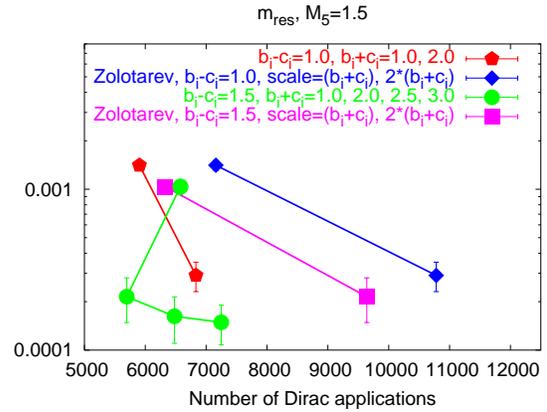}
\caption{Comparison of Zolotarev polynomial with polar
decomposition. Scale = $(b_i+c_i)$ stands for a choice of the
coefficients such that the Zolotarev polynomial, with $L_s=10$,
overlaps with the polar decomposition at $L_s=16$ as well as
possible (obviously, there is no overlap in the interval where the
Zolotarev polynomial is flat).}
\label{fig:Fig6}
\vskip-.174cm
\end{figure}

\subsection{Zolotarev coefficients and figure \ref{fig:Fig6}}

As mentioned above, the scaling coefficients $b_i+c_i$ can take $L_s$
different values. The optimal choice for the approximation to the sign
function are the Zolotarev coefficients \cite{Chiu:2002ir}. In other
words, of all possible choices for the coefficients, Zolotarev will
achieve the smallest residual mass. In fig.(\ref{fig:Fig6}) we show
results that employ Zolotarev's coefficients at $L_s=10$. We compare
with the polar decomposition (i.e.\ all coefficients equal) at
$L_s=16$. The $L_s$ are chosen such that the two polynomials overlap
as well as possible. The graph illustrates that Zolotarev's
performance is worse than what we found for the polar
decomposition. Even though Zolotarev's $L_s$ is much smaller, its
number of iterations in the linear system solver explodes.

This surprising behaviour is due to the fact that the convergence of
the linear system solver degrades with increasing $b_i+c_i$. For
Zolotarev there are always coefficients that are larger than the ones
being used for the polar decomposition. Even though there are smaller
ones as well, the large ones are responsible for the slow convergence.

\bibliographystyle{h-elsevier}
\bibliography{mobius}

\end{document}